\documentclass[12pt]{article}

\newcommand{\m}{\mathrm}
\newcommand{\be}{\begin{equation}}
\newcommand{\ee}{\end{equation}}
\newcommand{\ba}{\begin{eqnarray}}
\newcommand{\ea}{\end{eqnarray}}

\usepackage{graphicx}
\usepackage{amssymb}
\usepackage{amsmath}
\usepackage[T1]{fontenc} 
\usepackage[ansinew]{inputenc} 
\usepackage[nosort]{cite}
\newcommand{\inbar}{\vrule height1.57ex width.4pt depth0pt}
\newcommand{\SW}{\relax{\hbox{$\ \inbar\kern-.285em{\rm S}$}}}

\begin{document}
\thispagestyle{empty}
\begin{center}

\null \vskip-1truecm \vskip2truecm

{\Large{\bf \textsf{Large Numbers in Holography}}}

{\large{\bf \textsf{}}}

{\large{\bf \textsf{}}}

\vskip1truecm

{\large \textsf{Brett McInnes}}

\vskip1truecm

\textsf{\\  National
  University of Singapore}

\textsf{email: matmcinn@nus.edu.sg}\\

\end{center}
\vskip1truecm \centerline{\textsf{ABSTRACT}} \baselineskip=15pt
\medskip
The AdS/CFT correspondence is useful primarily when the number of colours, $N_{\textsf{c}}$, characterising the boundary field theory, is ``large'', and when the mass of the bulk black hole that is usually present is ``large'' relative to the bulk Planck mass. But this prompts two questions: first, can these large numbers be estimated, even very approximately, in a given application? Second: if these quantities are themselves computed holographically from physical data constraining the field theory, is this computation self-consistent, in the sense that it actually produces large numbers ---$\,$ an outcome which is far from obvious? Here we consider these questions in the case of the application of holographic techniques to the study of the quark-gluon plasma. We find that holography in this case is able to generate estimates of the dimensionless numbers in question, and, very remarkably, they are indeed large, despite the fact that the dimensionless input data are of order unity.

\newpage

\addtocounter{section}{1}
\section* {\large{\textsf{1. The Central Role of Large Numbers in Applied Holography}}}
The ``holographic dictionary'' \cite{kn:casa,kn:nat,kn:bag} is at the heart of the AdS/CFT correspondence and its applications. It allows us to translate the language of bulk physics into that of the boundary field theory, as follows (for a five-dimensional bulk spacetime, the only case to be considered in this work):
\begin{equation}\label{A}
\ell_{\textsf{P}}^3\; =\; {\pi\over 2}\times L^3 \times {1\over N_{\textsf{c}}^2},\;\;\;\;\;\;\;\ell_{\textsf{s}}^4\;=\;p \times L^4 \times {1\over \lambda},
\end{equation}
where $N_{\textsf{c}}$ is the number of colours in the boundary field theory, $\lambda$ is the 't Hooft coupling in that theory, $p$ is a certain pure number determined by the normalization of the gauge algebra basis\footnote{In the conventions of \cite{kn:nat}, $p = 1$; other conventions are possible, but of course $p$ is always of that order.}, $\ell_{\textsf{s}}$ is the string length scale, $\ell_{\textsf{P}}$ is the bulk Planck length, and $L$ is the asymptotic AdS$_5$ curvature length scale\footnote{Throughout this work, we use natural units, with either the electron volt or the femtometre (1 fm $\approx (197.3$ MeV)$^{-1}$) as the base; we never use Planck units.}.

We have written these equations in this form to emphasise their structure: on the left one has some physical parameter in the bulk, while the right side takes the form ``universal pure number, times a dimensionally appropriate power of $L$, times (some function of) a quantity characterising the boundary field theory''. 

Written in this way, the dictionary shows the fundamental role of $L$, as the quantity that mediates between bulk and boundary parameters; note in this connection that $L$ may be regarded as inhabiting both the bulk \emph{and} the boundary (that is, it influences both geometries, a fact that is particularly clear when the bulk contains a black hole with a topologically spherical event horizon). Note too that $L$ can in principle (of course, not in practice) be determined directly by measuring lengths and angles in the asymptotic bulk; it is not freely disposable.

The duality is useful primarily when the bulk geometry can be treated (semi-)classically and when strings propagating there can be treated approximately as point particles. A necessary condition for the first requirement is that the asymptotic geometric parameter $L$ should in some sense be large relative to $\ell_{\textsf{P}}$. The first entry in (\ref{A}) then allows us to interpret this demand as the well-known condition that $N_{\textsf{c}}$ should be ``large''. The second requirement, that it should be possible to approximate bulk physics by a theory of point particles, means that $L$ must be large relative to $\ell_{\textsf{s}}$; this leads to the condition that $\lambda$ must be ``large'', so that the boundary theory must be \emph{strongly coupled}. Conversely, we stress (because it will be relevant, below) that holographic models treating strings as point particles are \emph{not} applicable to situations where the boundary field theory becomes sufficiently weakly coupled.

In most applications, there is actually a third member of the holographic dictionary. This arises when the boundary theory is thermal: in that case, there must be a semi-classical black hole in the bulk, with a Hawking temperature dual to the temperature of the boundary matter. In practice, computations using this black hole rely in a crucial way on the details of the classical geometry, so it is essential that quantum-gravitational effects in this geometry should be suppressed; this evidently does not follow from the requirement, discussed above, that the \emph{asymptotic} region should have an approximately classical geometry. That is, if $m_{\textsf{P}} = 1/\ell_{\textsf{P}}$ is the AdS$_5$ Planck mass, we must ensure that $\mathcal{M}/m_{\textsf{P}}$, where $\mathcal{M}$ is the physical mass of the black hole, is a (very) large number. The relevant entry in the dictionary will be derived below: expressing $\mathcal{M}/m_{\textsf{P}}$ as $\mathcal{M}/m_{\textsf{P}} = \mathcal{M}\ell_{\textsf{P}}$ (we will do this henceforth), it has the form, for an AdS$_5$-Kerr black hole,
\begin{equation}\label{AARDVARK}
{\mathcal{M}\ell_{\textsf{P}}\over M}\,{\left[1 - \left(a/L\right)^2\right]^2\over 3 - \left(a/L\right)^2} \;=\; \left({\pi \over 16}\right)^{{1\over 3}}\times L^{- 2} \times N_{\textsf{c}}^{4/3},
\end{equation}
where $M$ and $a$ are certain parameters, to be defined below, describing the black hole geometry. (Note that $M$ has units of length squared, that is, despite the notation, it is not a mass.) From this we see that (unless $N_{\textsf{c}}$ is \emph{extremely} large, or $a$ is very close to $L$, neither of which turns out to be the case) the requirement that the black hole be effectively classical means that the dimensionless combination $M/L^2$ should be very large ---$\,$ \emph{despite the fact that $L$ itself is large}.

One often speaks as if the word ``large'' in these statements meant ``arbitrarily large'', but of course this is not so: in particular\footnote{In this work, we will consider only the magnitudes of $N_{\textsf{c}}$ and $\mathcal{M}\ell_{\textsf{P}}$. Different techniques are required to answer analogous questions regarding $\lambda$.}, the finiteness of $N_{\textsf{c}}$ plays an important role in several applications.

For example, asymptotically AdS black holes are claimed to be ``fast scramblers'' \cite{kn:sus}; this is connected with chaotic behaviour in the boundary theory \cite{kn:mal,kn:arnab}. However, the scrambling time is an increasing function of some measure of the number of degrees of freedom describing the system, and in this case that number is related to $N_{\textsf{c}}$; thus, in order to understand the scrambling time more precisely, one needs some information regarding the (necessarily finite) value of $N_{\textsf{c}}$.

Again, the ``$1/N_{\textsf{c}}$ corrections'', arising when $N_{\textsf{c}}$ is taken to be large but not so large that $1/N_{\textsf{c}}$ is altogether negligible, are very important: to take one example of many, it has been suggested that these corrections may explain certain phenomena connected with energy-momentum transport in strongly coupled fluids \cite{kn:oka,kn:wolf}. Here, too, it seems quite possible that it would be very helpful if one could be more specific: how large counts as ``large'' $N_{\textsf{c}}$? Even a very approximate or semi-quantitative answer to this basic question could be useful, particularly since $1/N_{\textsf{c}}$ corrections correspond to string loop (that is, quantum-gravitational) effects in the bulk, and so it is very difficult to perform explicit calculations.

It is worth mentioning in this connection that, in lattice gauge theory, where of course large numbers of colours are often considered, ``large'' $N$ typically means up to about 20 \cite{kn:gun}; recent advances have allowed consideration of values in the hundreds \cite{kn:marg}, but not larger. This gives another perspective on the meaning of ``large'' $N_{\textsf{c}}$ in holography.

In summary, then: in a given application of holography, $N_{\textsf{c}}$ has a large but finite value. One would like to know something about that value.

Now, in applications, the properties of the bulk and boundary physics are constrained by the physics of the system that the boundary field theory approximates. For example, a black hole in the bulk has a certain entropy, and this is related holographically to the entropy density of the boundary field theory; and this in turn should be related to the entropy density of the physical system being approximated by the field theory. The simple but crucial observation we make here is this: with sufficient data from the physics we hope to study holographically, one might be able to use holography itself to compute or at least constrain some of the parameters in the bulk/boundary system, \emph{including} perhaps $L$ and $N_{\textsf{c}}$. We will show that this can indeed be done.

This is interesting in itself. \emph{But, in addition, it presents the entire programme with a fundamental challenge}. For it is by no means clear that the numbers resulting from such a calculation will be in any sense ``large'': on the contrary, as we will show, the dimensionless input parameters are often of order unity, and one consequently has no reason to expect large numbers in the output.

For example, we stressed earlier that it is essential that the black hole mass $\mathcal{M}$ should be such as to put the bulk geometry firmly in the classical domain: that is, we need $M/L^2$, which can be computed holographically given sufficient boundary data, to be a very large number. There is little hope that such a large number can arise from input which involves no such quantities. \emph{There appears to be a real risk of internal inconsistency here}.

In this work we consider these questions in a specific context: the regime \cite{kn:nat} in which the boundary field theory in the AdS/CFT correspondence bears some resemblance to the Quark-Gluon Plasmas (QGP) produced in heavy-ion collisions \cite{kn:bus}. We stress that our results apply only to this concrete example; but one can hope that the lessons learned in this case may be useful more generally.

We obtain explicit order-of-magnitude estimates of $N_{\textsf{c}}$ and $\mathcal{M}\ell_{\textsf{P}}$. We find, against all odds, that \emph{the equations of the holographic model of the thermal QGP do in fact have numerically ``large'' solutions}: for example, even with dimensionless input data of order unity, the computed value for $\mathcal{M}\ell_{\textsf{P}}$ (using data from central heavy ion collisions at $27$ GeV per pair) is of order $10^{19}$, so that the use of a semi-classical black hole geometry for the bulk is consistent.

The argument runs as follows.

\addtocounter{section}{1}
\section* {\large{\textsf{2. A Strategy for Computing the Large Numbers}}}
In the case of peripheral heavy-ion collisions, it is well known that the plasma often acquires a substantial angular momentum density; this effect has actually been (indirectly) observed by the STAR collaboration at the RHIC facility \cite{kn:STAR,kn:STARcoll,kn:STARcoll2,kn:tan,kn:zuo}. The dual system is then a five-dimensional rotating black hole with a non-zero angular momentum to mass ratio $\mathcal{A}$. This quantity is dual to the ratio of the angular momentum density of the plasma to its energy density, quantities which can be estimated in phenomenological models such as those described in \cite{kn:sahoo,kn:jiang}. This could be useful, as follows.

In the study of such black holes, and in particular when the question of Cosmic Censorship is considered \cite{kn:96,kn:98}, one finds\footnote{To be precise, we do not have a proof of this statement valid in all cases. But in \cite{kn:98} it is shown that it \emph{is} valid for AdS$_5$ black holes with parameters such that the boundary field theory has the same temperature, energy density, and so on, as the actual QGP; and these are the black holes with which we are concerned in this work.} that $\mathcal{A}$ satisfies a very remarkable inequality: we have
\begin{equation}\label{B}
\mathcal{A}\;<\;L,
\end{equation}
where $L$ is the asymptotic AdS$_5$ length scale, as above. This restriction is at first sight very surprising, since it implies a strong link between the immediate vicinity of the black hole, where $\mathcal{A}$ provides the dominant length scale, and the asymptotic region, where the geometry is primarily controlled by $L$. It can be partially understood from the fact that, unlike asymptotically flat rotating black holes, asymptotically AdS black holes give rise to a \emph{frame-dragging} effect which does not decay to zero with distance from the black hole. In this way, $\mathcal{A}$ still makes its presence felt in the asymptotic region.

The fact that $\mathcal{A}$ is bounded by (a multiple of) $L$ is explained by string theory \cite{kn:96}: it turns out that, in that context, $\mathcal{A}$ and $L$ \emph{compete} to determine the sign of the asymptotic action of certain branes, with $\mathcal{A}$ tending to force it to be unbounded below. Thus, the system becomes severely unstable if $\mathcal{A}$ is too large relative to $L$. (The statement that the coefficient of $L$ on the right side of (\ref{B}) is precisely unity has to be established by further argument: see \cite{kn:98}.)

The statement that $\mathcal{A}$ is bounded above by $L$ is relevant to us here, because it can be reinterpreted as the statement that $L$ is bounded \emph{below} by $\mathcal{A}$, and, as explained above, holographic duality allows us to give a rough approximate value for the latter by referring to experimental data; so we can use such data, for RHIC experiments producing plasmas with maximal angular momenta, to put a lower bound on $L$ in this specific application. In fact, with more data, we can do better: we can actually \emph{estimate} $L$ for RHIC plasmas using the holographic model of the vortical QGP. Crucially, we will see that this can be done \emph{without} knowing either $N_{\textsf{c}}$ or $\ell_{\textsf{P}}$.

Having an actual, if admittedly very approximate, value for $L$ in this concrete example, has two advantages. First, it allows us to determine whether $L$ is ``large'', not just in the abstract context of a discussion of the status of the holographic dictionary ---$\,$ though of course that is vitally important for the method to be internally consistent ---$\,$ but also in the sense of being large relative to the actual physical length scales describing the QGP. The RHIC plasmas \cite{kn:STAR} are characterised by length scales defined, for example, by their temperature, baryonic chemical potential, and so on. Typically these are on the order of a femtometre, and so ``large'', in this specific case, means that the value we compute for $L$ should be significantly larger than length scales of that order. This gives us a second consistency check.

Secondly, and more importantly, having a definite value for $L$ allows us to estimate $N_{\textsf{c}}$, as follows.

The holographic model described above uses the uncharged AdS$_5$-Kerr black hole geometry, so \cite{kn:casa} it does not account for the \emph{baryonic chemical potential} of the QGP. Therefore, it only applies directly to relatively high impact energies, such that the resulting plasma has a negligible baryonic chemical potential relative to the temperature; that is, particles and antiparticles are produced in approximately equal numbers. At lower impact energies, the baryonic chemical potential is not negligible compared to the temperature, so we need to endow the black hole with charge. For this purpose, we simplify matters by focusing on \emph{central} collisions, so that now the angular momentum can be neglected. This means that we reduce the angular momentum of the bulk black hole to a negligible value, and then increase the electric charge from zero up to the required value\footnote{This clumsy procedure is imposed on us by the fact that the metrics of asymptotically AdS$_5$ black holes endowed with \emph{both} angular momentum and charge ---$\,$ that is, AdS$_5$-Kerr-Newman black holes ---$\,$ are not currently known \cite{kn:emparan}.}. The black hole is then of the AdS$_5$-Reissner-Nordstr\"om variety (with a spherical, not planar, event horizon).

We will see that the holographic dictionary extends to this case: the new entry relates bulk parameters (including the electric charge parameter in the black hole metric) to boundary parameters, including the baryonic chemical potential, but \emph{also} including $N_{\textsf{c}}$. Again using phenomenological models of the RHIC plasmas produced in central collisions (together with the value of $L$ computed earlier) we can therefore use the dictionary to estimate the value of $N_{\textsf{c}}$ appropriate to a holographic model of this specific system (plasmas produced in central collisions at relatively low impact energies).

Here we will carry out a basic version of this programme. The objective at this point is not to evaluate $L$ or $N_{\textsf{c}}$ precisely, but rather to determine whether these fundamental parameters are in fact ``large''. We will refine this a little: if indeed they \emph{are} large, are they merely ``moderately large'' or instead ``extremely large''? For definiteness, we will take ``moderately large'' to mean, in the case of $N_{\textsf{c}}$, roughly one to two orders of magnitude larger than 3, the QCD value; for $L$ we will take it to mean roughly one to two orders of magnitude larger than the characteristic length scale defined by QCD, that is, about a femtometre (corresponding to the QCD energy scale, around 200 MeV). Even this simple and admittedly somewhat arbitrary distinction could be important: it could tell us, for example, whether the $1/N_{\textsf{c}}$ corrections are in fact likely to be totally negligible in \emph{all} applications of holographic duality to QGP physics, and perhaps even whether there is any realistic hope of ultimately connecting large-$N_{\textsf{c}}$ theories with QCD. The point is simply that $N_{\textsf{c}} = 30$ and $N_{\textsf{c}} = 3000$ are both ``large'', but their consequences could be very different in some cases.

We begin with a brief review of the main results of \cite{kn:98}, which involve the holography of (singly-rotating) AdS$_5$-Kerr black holes, and use them to compute an approximate value for $L$ in this case. In the subsequent Section we use this value to estimate $N_{\textsf{c}}$.

\addtocounter{section}{1}
\section* {\large{\textsf{3. Using AdS$_5$-Kerr Black Holes to Estimate $L$}}}
The singly-rotating\footnote{In general, an AdS$_5$-Kerr metric is described by two rotation parameters, $(a, b)$ \cite{kn:hawk}. Here we set $b = 0$, hence the notation $(a, 0)$.} AdS$_5$-Kerr metric \cite{kn:hawk,kn:cognola,kn:gibperry}, expressed in terms of Hopf coordinates on the three-sphere, is
\begin{flalign}\label{C}
g\left(\m{AdSK}_5^{(a,0)}\right)\; = \; &- {\Delta_r \over \rho^2}\left[\,\m{d}t \; - \; {a \over \Xi}\,\m{sin}^2\theta \,\m{d}\phi\right]^2\;+\;{\rho^2 \over \Delta_r}\m{d}r^2\;+\;{\rho^2 \over \Delta_{\theta}}\m{d}\theta^2 \\ \notag \,\,\,\,&+\;{\m{sin}^2\theta \,\Delta_{\theta} \over \rho^2}\left[a\,\m{d}t \; - \;{r^2\,+\,a^2 \over \Xi}\,\m{d}\phi\right]^2 \;+\;r^2\cos^2\theta \,\m{d}\psi^2 ,
\end{flalign}
where
\begin{eqnarray}\label{D}
\rho^2& = & r^2\;+\;a^2\cos^2\theta, \nonumber\\
\Delta_r & = & \left(r^2+a^2\right)\left(1 + {r^2\over L^2}\right) - 2M,\nonumber\\
\Delta_{\theta}& = & 1 - {a^2\over L^2} \, \cos^2\theta, \nonumber\\
\Xi & = & 1 - {a^2\over L^2}.
\end{eqnarray}
Here $M$ is a geometric parameter (with units of squared length) which is related but not equal to the physical mass of the black hole, $\mathcal{M}$: the latter is
\begin{equation}\label{DRONGO}
\mathcal{M}\;=\;{\pi M \left(2 + \Xi\right)\over 4\,\ell_{\textsf{P}}^3\,\Xi^2}.
\end{equation}
Using the holographic dictionary, one can express this more usefully as
\begin{equation}\label{DAGGY}
\mathcal{M} \ell_{\textsf{P}} \;=\; \left({\pi \over 16}\right)^{{1\over 3}}\,{2 + \Xi\over \Xi^2}\,N_{\textsf{c}}^{4/3}\,{M\over L^2},
\end{equation}
and this, with some rearrangement, is the third entry in the holographic dictionary, equation (\ref{AARDVARK}) above.

Similarly, $a$ is a geometric parameter (with units of length) which ranges between zero and $\sqrt{3}\,L$ (excluding $L$ itself), which is \emph{not} equal to the (physical) angular momentum per (physical) mass $\mathcal{A}$; but, together with $L$, it does determine that quantity through the relation \cite{kn:gibperry}
\begin{equation}\label{E}
\mathcal{A}\;=\;{2 a \over 2 + \Xi}\;=\;{2 a \over 3 - \left(a^2/L^2\right)}.
\end{equation}
Note that $\mathcal{A}/L < 1$ if and only if $a/L < 1$; note also, however, that $\mathcal{A}$ has a very different range to that of $a$, and can, in fact, be arbitrarily large (without necessarily violating Cosmic Censorship, see \cite{kn:96}).

The product of the black hole specific entropy $\mathcal{S}$ (entropy per unit physical mass) with the Hawking temperature, $T_{\textsf{H}}$, is a very useful\footnote{$\mathcal{S}T_{\textsf{H}}$ and $\mathcal{A}T_{\textsf{H}}$ are useful because they have no explicit dependence on $\ell_{\textsf{P}}$, and they depend on $L$ only through the dimensionless combinations $a/L, M/L^2$, and $r_{\textsf{H}}/L$.} dimensionless quantity, given in this case by
\begin{equation}\label{UNO}
\mathcal{S}T_{\textsf{H}}\;=\;{2\,|\Xi|\over 2 + \Xi}\,r_{\textsf{H}}^2\,\left({1\over r_{\textsf{H}}^2 + a^2}\,+\,{1\over r_{\textsf{H}}^2 + L^2}\right).
\end{equation}
We will also find it useful to consider the dimensionless (in natural units) quantity $\mathcal{A}T_H$, which takes the form
\begin{equation}\label{DUE}
\mathcal{A}T_{\textsf{H}}\;=\;{a r_{\textsf{H}} \over \pi \left(2 + \Xi\right)}\,\left({1 + {r_{\textsf{H}}^2\over L^2}\over r_{\textsf{H}}^2 + a^2} + {1\over L^2}\right).
\end{equation}
In both these equations, $r_{\textsf{H}}$ is the horizon ``radius''; its dimensionless version, $r_{\textsf{H}}/L$, is given as a function of the dimensionless parameters $M/L^2$ and $a/L$ by solving the equation
\begin{equation}\label{F}
\left({r_{\textsf{H}}^2\over L^2}+{a^2\over L^2}\right)\left(1 + {r_{\textsf{H}}^2\over L^2}\right) - 2\,{M\over L^2}\;=\;0.
\end{equation}
We mention in passing that the derivations of (\ref{UNO}) and (\ref{DUE}) do not depend on assuming $a < L$; it is possible in some cases for (\ref{F}) to have real solutions, which can be inserted into (\ref{UNO}) and (\ref{DUE}) (that is, for classical Censorship to hold), when this condition is not satisfied (see \cite{kn:96,kn:98} for extensive discussions of this point). However, in the present work, $a/L$ (and therefore $\mathcal{A}/L$) is always smaller than unity.

Given experimental (in practice, phenomenological) values for the temperature, and for the angular momentum, energy, and entropy densities of the QGP produced in heavy-ion collisions at a given impact energy and centrality, holography fixes the left sides of (\ref{UNO}) and (\ref{DUE}) in a straightforward way (for example, $\mathcal{S}$ is dual to the ratio of the entropy and energy densities of the boundary matter); and so one can solve those equations, with (\ref{F}), for the three unknowns $a/L, M/L^2$, and $r_{\textsf{H}}/L$.

It is important to note, however, that this holographic model is adequate only to the extent that the baryonic chemical potential $\mu_{\textsf{B}}$ is negligible compared to the temperature, since the black hole must have electric charge in order to represent non-zero $\mu_{\textsf{B}}$ \cite{kn:casa}, and these black holes are electrically neutral. In this section, therefore, we consider only plasmas produced in the highest impact energy collisions at the RHIC facility\footnote{We have not used data from the heavy-ion collisions at the LHC; for it appears that holographic methods are less reliable in this case than in the case of RHIC plasmas. To see the point, note that the ratio of the shear viscosity to entropy density, $\eta/s$, for the 200 GeV collisions studied at the RHIC, is not far above the well-known holographic bound, $1/4\pi$; but, for the plasmas produced in 2.76 TeV collisions at the LHC, the value of $\eta/s$ may be \emph{up to $60\%$ larger} \cite{kn:malg}. This suggests that, while a hydrodynamic description of the post-collision system is still appropriate at LHC collision energies, the holographic implementation of that description may be significantly less reliable in this case. Possibly this is due to the plasma being less strongly coupled in the LHC case (see our discussion of the role of strong coupling in applied holography, above).}, that is, 200 GeV per pair; the baryonic chemical potential, about 27 MeV \cite{kn:STARchem}, is indeed small compared with the temperature in this case (around 190 MeV). We also focus on the centrality (about $17\%$) that maximizes the angular momentum density.

Using the phenomenological models described in \cite{kn:sahoo,kn:jiang}, one finds \cite{kn:93,kn:95} that, for collisions at $\sqrt{s_{NN}} = 200$ GeV and centrality $17\%$, the ratio of the angular momentum density to the energy density, which corresponds holographically to $\mathcal{A}$, is about $77$ femtometres. It follows from the inequality (\ref{B}) that $L$ \emph{must be at least this large}; this in fact already means that $L$ is ``large'' relative to the obvious length scales associated with this system (for example, the inverse temperature is just over one femtometre).

In fact, we can do better than this, by using additional data from \cite{kn:sahoo}. There we find, for example, that the ratio of the entropy density to the energy density for plasmas produced in 200 GeV collisions is about $6.5 \times 10^{-3}$ MeV$^{- 1}$ , and the temperature is about $190$ MeV. As above, the maximal value of $\mathcal{A}$, the ratio of the angular momentum and energy densities, is around $77$ fm (or $0.39$ MeV$^{- 1}$). As discussed earlier, holography now allows us to use these data to compute the left sides of equations (\ref{UNO}) and (\ref{DUE}); combining the resulting equations with equation (\ref{F}), we can now determine $a/L$ and $M/L^2$. We find that $a/L\approx 0.322$ in this case. Using equation (\ref{E}), we now obtain $\mathcal{A}/L \approx 0.222,$ and so $\mathcal{A} \approx 77$ fm gives us finally $L \approx 350$ fm.

We see that $L$, in this specific application, is considerably larger than the inequality (\ref{B}) requires; and it is even larger relative to the usual length scales of QGP physics. We conclude that there is no internal inconsistency in applying holographic methods to this system: $L$ is in fact what we agreed to call ``moderately large''.

Next, we consider the value of $M/L^2$ that emerges from this calculation. It is, in view of the modest input values (and the rather simple equations involved), rather startling: we find $M/L^2 \approx 6.06\times10^{11}$. In the next section, we will show that $N_{\textsf{c}}$ is roughly 30 in these simple models: inserting this and $a/L\approx 0.322$ into (\ref{DAGGY}), we find that $\mathcal{M}\ell_{\textsf{P}} \approx 1.18 \times 10^{14}$. This suffices to ensure that the classical approximation for this black hole is reasonably good.

In summary: our use of classical black hole geometry in the bulk is fully consistent: \emph{holography itself gives rise to the requisite very large number}.

We note in passing that $L \approx 350$ fm means that the volume of the spatial sections at infinity (with a suitable choice of the conformal factor) is several tens of millions of cubic femtometres, vastly larger than the actual initial volume of the plasma produced in a heavy-ion collision (about 100 fm$^3$); so the spatial sections are effectively flat.

We are now in a position to estimate $N_{\textsf{c}}$.

\addtocounter{section}{1}
\section* {\large{\textsf{4. Using AdS$_5$-Reissner-Nordstr\"om Black Holes to Estimate $N_{\textsf{c}}$}}}
As explained earlier, we now consider central collisions at lower impact energies. This means that we can set the angular momentum of the bulk black hole to zero, but now we must take its charge to be non-zero, so as to model the non-negligible baryonic chemical potential arising in the corresponding plasmas. That is, we need to consider AdS$_5$-Reissner-Nordstr\"om black holes.

An AdS$_5$-Reissner-Nordstr\"om black hole metric takes the form (for a spherical event horizon)
\begin{flalign}\label{G}
g(\m{RNAdS}_5)\;=\;&-\,\left({r^2\over L^2}\,+\,1\,-\,{2M\over r^2}\,+\,{Q^2\over 4\pi r^4}\right)\m{d}t^2\,+{\m{d}r^2\over {r^2\over L^2}\,+\,1\,-\,{2M\over r^2}\,+\,{Q^2\over 4\pi r^4}}\\ \notag \,\,\,\,&\,+\,r^2\left(\m{d}\theta^2 \,+\, \sin^2\theta\,\m{d}\phi^2\,+\,\cos^2\theta\,\m{d}\psi^2\right).
\end{flalign}
Here the angular coordinates are as in the preceding Section. The parameters $M$ and $Q$, which both have units of squared length, are connected with the physical mass and charge of the black hole, $\mathcal{M}$ and $\mathcal{Q}$, through the relations
\begin{equation}\label{H}
\mathcal{M}\;=\;{3\pi M\over 4\ell^3_{\textsf{P}}},\;\;\;\;\;\mathcal{Q}\;=\;{\sqrt{3}\pi Q\over 2\ell^{3/2}_{\textsf{P}}\sqrt{k_5}},
\end{equation}
where $\ell_{\textsf{P}}$ is the AdS$_5$ Planck length, as above, and where $k_5$ is the five-dimensional Coulomb constant. Using the holographic dictionary (or setting $a = 0$ in (\ref{DAGGY})), one also has
\begin{equation}\label{HORROR}
\mathcal{M}\ell_{\textsf{P}}\; =\;3\, \left({\pi \over 16}\right)^{{1\over 3}}\,N_{\textsf{c}}^{4/3}\,{M\over L^2}.
\end{equation}

Let us digress briefly to discuss the parameter\footnote{See \cite{kn:myers}, Section 2, for a different discussion of the fact that, in effect, an additional length scale enters holography at this point.} $k_5$.  Our convention in this work is that five-dimensional electric charge is dimensionless in natural units, as is the case in four dimensions. This means that $k_5$, unlike its four-dimensional counterpart, has dimensions of length or inverse energy; in fact, $1/k_5$ is the energy associated, in five dimensions, with two unit charges in the bulk, separated by $\delta r = k_5$. Holographically, then, one might think of $1/k_5$ as the characteristic energy scale of the dual system. In our case, that scale is the QCD scale, usually taken to be around 200 MeV; that is, $k_5$ is about one femtometre. When we interpret $1/k_5$ in this holographic manner, we will denote it by $E_{\textsf{QCD}}$.

The Hawking temperature of the AdS$_5$-Reissner-Nordstr\"om black hole is
\begin{equation}\label{I}
4\pi T_{\textsf{H}}\;=\;{8M\over r_{\textsf{H}}^3}\,-\,{2\over r_{\textsf{H}}}\,-\,{3Q^2\over 2\pi r_{\textsf{H}}^5},
\end{equation}
where $r_{\textsf{H}}$ corresponds to the outer event horizon; it can be regarded as a function of $M$, $Q$, and $L$ by means of the equation
\begin{equation}\label{J}
{r_{\textsf{H}}^2\over L^2}\,+\,1\,-\,{2M\over r_{\textsf{H}}^2}\,+\,{Q^2\over 4\pi r_{\textsf{H}}^4}\;=\;0.
\end{equation}
The black hole entropy per unit physical mass, $\mathcal{S}$, is
\begin{equation}\label{K}
\mathcal{S}\;=\;{2\pi r_{\textsf{H}}^3\over 3 M};
\end{equation}
notice that it does not depend on $\ell_{\textsf{P}}$.

The electromagnetic potential form is
\begin{equation}\label{L}
\mathbb{A}\;=\;{\sqrt{3}Q \sqrt{k_5}\over 8\pi \ell^{3/2}_{\textsf{P}}}\left({1\over r^2}\,-\,{1\over r^2_H}\right)\m{d}t;
\end{equation}
notice that the coefficient of d$t$ has the correct units, inverse length or energy. (The expression involves the square root of $k_5$ because the electric charge on the black hole is \emph{not} $Q$, but rather $\mathcal{Q}$: see above.)

As in the preceding Section, holography identifies $T_{\textsf{H}}$ with the temperature of the boundary matter, and $\mathcal{S}$ with $s/\varepsilon$ (where $s$ is the entropy density, and $\varepsilon$ the energy density). Finally $\mu_{\textsf{B}}$, the baryonic chemical potential of the boundary matter, is dual \cite{kn:clifford} to $- 3$ times the asymptotic value of the coefficient of d$t$ in $\mathbb{A}$, that is,
\begin{equation}\label{M}
\mu_{\textsf{B}}\;=\;{3 \sqrt{3} Q \sqrt{k_5}\over 8 \pi r_{\textsf{H}}^2 \ell_{\textsf{P}}^{3/2}}.
\end{equation}

Eliminating $\ell_{\textsf{P}}$ from this relation by means of the holographic dictionary (\ref{A}), we have a fourth entry in that dictionary:
\begin{equation}\label{N}
{Q\over r_{\textsf{H}}^2} = {4\sqrt{2}\pi^{3/2}\over 3\sqrt{3}} \times L^{3/2} \times {\mu_{\textsf{B}}\over N_{\textsf{c}}}\,\sqrt{E_{\textsf{QCD}}}.
\end{equation}
Here the quantities on the left are, as in (\ref{A}), bulk parameters, linked to boundary theory parameters by a power of $L$, which we now know. (As explained above, we interpret $1/k_5$ holographically as a characteristic energy scale for the boundary matter, $E_{\textsf{QCD}}$, so this too is known.)

Using again the phenomenological values for temperature, entropy density, and energy density from \cite{kn:sahoo}, and the observed baryonic chemical potential from \cite{kn:STARchem}, we can fix $T_{\textsf{H}}$, $\mathcal{S}$, and $\mu_{\textsf{B}}$ in equations (\ref{I}), (\ref{J}), (\ref{K}), and (\ref{N}), and, using also $L \approx 350$ fm, solve them for the four unknowns, $M$, $Q$, $r_{\textsf{H}}$, and, in particular, $N_{\textsf{c}}$.

We need to do this for an impact energy which, on the one hand, is not so low that it is questionable whether a QGP is formed at all, but which, on the other, is not so large that $\mu_{\textsf{B}}$ is negligible. We will therefore consider RHIC impact energy $\sqrt{s_{\m{NN}}} = 27$ GeV per pair, and use the corresponding data from \cite{kn:sahoo} and \cite{kn:STARchem}, as above. The initial temperature in this case is around $172$ MeV, the ratio of entropy and energy densities, $\mathcal{S}$, is around $6.8 \times 10^{-3}$ MeV$^{-1}$, and $\mu_{\textsf{B}}$ is about $160$ MeV. Other RHIC impact energies in this intermediate range (such as 19.6 and 39 GeV) lead to results which differ somewhat from those obtained below; but they are of the same respective orders of magnitude.

Solving the equations with these input data, one finds that $N_{\textsf{c}} \approx 30$ (``moderately large''), while $M/L^2 \approx 6.8 \times 10^{16}$. Inserting these numbers into equation (\ref{HORROR}), we find that the ratio of the black hole mass to the bulk Planck mass is $\mathcal{M}\ell_{\textsf{P}} \approx 1.14 \times 10^{19}$.

Clearly, the apparently simple set of equations (\ref{I}), (\ref{J}), (\ref{K}), (\ref{N}) has some unusual mathematical properties, and we hope to return to this elsewhere. In the meantime, we can summarize as follows:

$\bullet$ \emph{Bearing in mind that this is no more than an order-of-magnitude estimate}, $N_{\textsf{c}}$ takes a value here that is [a] similar to the values for $N$ often used in large-$N$ lattice computations, [b] arguably large enough to justify the application of holographic methods in this case, but [c] not so large that the $1/N_{\textsf{c}}$ corrections \cite{kn:oka,kn:wolf} are completely negligible under all circumstances.

$\bullet$ The black hole mass $\mathcal{M}$ in this case exceeds the bulk Planck mass by a factor which ensures that the black hole can indeed be consistently treated approximately classically.

\addtocounter{section}{1}
\section* {\large{\textsf{5. Conclusion}}}
It is a familiar limitation \cite{kn:mateos} of ``applied holography'' that the method works best when certain numbers are ``large''. It is natural to ask: how large is ``large''? In this work, we have attempted to provide a preliminary, order-of-magnitude answer to such questions: for example, $N_{\textsf{c}}$ is probably large enough to justify neglecting $1/N_{\textsf{c}}$ corrections in most cases, but not in all.

The argument prompts another observation: there was no reason to expect the outcomes of such calculations to be ``large''. On the contrary, there is every reason to expect holography to fail this consistency test. For example, in the case of central heavy-ion collisions at 27 GeV per pair, the dimensionless parameters $\mathcal{S}T_{\textsf{H}} \approx 1.17$ and $\mu_{\textsf{B}}/T_{\textsf{H}} \approx 0.93$ generated by the data are close to unity: how then can we expect numbers several (or many) orders of magnitude larger to emerge from such a simple holographic model as the one we are using here?

Yet they do: the model does predict large numbers where they are needed. This in itself gives some reassurance that those numbers are reliable; so they can then be used to assess the whole project. For example, one might begin to construct arguments in favour of the claim that, while $N_{\textsf{c}} \approx 30$ is far from QCD, it is still near enough to offer hope that holography might in future draw nearer to a more realistic account of the QGP.

\addtocounter{section}{1}
\section*{\large{\textsf{Acknowledgement}}}
The author is grateful to Dr. Soon Wanmei for useful discussions.

\end{document}